\documentclass[12pt,pra,aps,amssymb,amsmath,tightenlines]{revtex4}

\newcommand{\rd}{\mbox{$\rm d$}}

\newcommand{\R}{\mathbb{R}}

\newcommand{\E}{\mathbb{E}}
\newcommand{\N}{\mathbb{N}}
\newcommand{\F}{\mathcal{F}}
\newcommand{\e}{\textrm{e}}
\newcommand{\nn}{\nonumber}
\begin{document}
\title{Information, Inflation, and Interest}
\author{Lane~P.~Hughston and Andrea~Macrina}
\affiliation{Department of Mathematics, King's College London, The
Strand, London WC2R 2LS, UK}
%
%
\begin{abstract}
{\bf Abstract}. We propose a class of discrete-time stochastic
models for the pricing of inflation-linked assets. The paper
begins with an axiomatic scheme for asset pricing and interest
rate theory in a discrete-time setting. The first axiom introduces
a ``risk-free" asset, and the second axiom determines the
intertemporal pricing relations that hold for dividend-paying
assets. The nominal and real pricing kernels, in terms of which
the price index can be expressed, are then modelled by introducing
a Sidrauski-type utility function depending on (a) the aggregate
rate of consumption, and (b)  the aggregate rate of real liquidity
benefit conferred by the money supply. Consumption and money
supply policies are chosen such that the expected joint utility
obtained over a specified time horizon is maximised subject to a
budget constraint that takes into account the ``value" of the
liquidity benefit associated with the money supply. For any choice
of the bivariate utility function, the resulting model determines
a relation between the rate of consumption, the price level, and
the money supply. The model also produces explicit expressions for
the real and nominal pricing kernels, and hence establishes a
basis for the valuation of inflation-linked securities.
\\ \vspace{0.0cm}

\\ \vspace{0.0cm}

\noindent Working paper. This version: 14 October 2007.

\noindent E-mail: lane.hughston@kcl.ac.uk,
andrea.macrina@kcl.ac.uk
\end{abstract}
\maketitle
\section{Introduction}
In this paper we apply the information-based asset pricing scheme proposed in
Brody {\it et al.} 2006, 2007, and Macrina 2006, to introduce a class of discrete-time models for interest rates and
inflation. The key idea is that market participants have at any
time partial information about the future values of the macro-economic
factors that determine interest rates and price levels. We
present a model for such partial information, and show how it
leads to a novel framework for the arbitrage-free dynamics of
real and nominal interest rates, price-indices, and index-linked
securities.

We begin with a general model for discrete-time asset pricing. We
take a pricing kernel approach, which has builds
in the arbitrage-free property and provides the desired link to
economic equilibrium. We require that the pricing kernel should be
consistent with a pair of axioms, one giving the
intertemporal relations for dividend-paying assets, and the other
relating to the existence of a money market asset. Instead of
directly assuming the existence of a previsible money-market
account, we make a weaker assumption, namely the
existence of an asset that offers a positive rate of return. It
can be deduced, however, that the existence of a
positive-return asset is sufficient to imply the existence of a
previsible money-market account, once the intertemporal relations
implicit in the first axiom are taken into account.

The main result of Section II is the derivation of a general
expression for the price process of a limited-liability asset.
This expression includes two terms, one being the
discounted and risk-adjusted value of the dividend stream, and the
other characterising retained earnings. The vanishing of the
latter is shown in Proposition 1 to be given by a transversality
condition, equation (\ref{2.3}). In particular, we are able to
show (under the conditions of Axioms A and B) that in the case of
a limited-liability asset with no permanently retained earnings,
the general form of the price process is given by the ratio of a
potential and the pricing kernel, as expressed in equation
(\ref{2.13}). In Section III we consider the per-period rate of
return $\{\bar{r}_i\}$ offered by the positive return asset, and
we show in Proposition 2 that there exists a constant-value asset
with limited liability such that the associated dividend flow is
given by $\{\bar{r}_i\}$. This result is then used in Proposition
3 to establish that the pricing kernel admits a decomposition of
the form (\ref{3.7}). In Proposition 4 we prove a converse to this result, thus giving a
procedure for constructing examples of systems satisfying Axioms A
and B. The method involves the introduction of an increasing
sequence that converges to an integrable random variable. Given
the sequence we construct an associated pricing kernel and
positive-return asset satisfying the intertemporal relations.

In Section IV  we introduce the nominal discount bond system
arising with the specification of the pricing kernel, and in
Proposition 5 we show that the discount bond system admits a
representation of the Flesaker-Hughston type. In Section V we
consider the case when the positive-return asset has a previsible
price process, and hence can be interpreted (in a
standard way) as a money-market account, or ``risk-free" asset.
The results of the previous sections do not depend on this
additional assumption. A previsible money-market account has the
structure of a series of one-period discount-bond investments.
Then in Proposition 6 we show under the conditions of Axioms A and
B that there exists a unique previsible money-market account. In
other words, although we only assume the existence of a
positive-return asset, we can establish the existence of a
money-market account.

In Section VI we outline a general approach to interest rate
modelling in the information-based framework, in a discrete-time setting.
In Section VII we then propose a class of models for the
pricing of inflation-linked assets. The nominal and real pricing
kernels, in terms of which the consumer price index can be
expressed, are modelled by introducing a bivariate utility
function depending on (a) consumption, and (b) the
real liquidity benefit conferred by the money supply.
Consumption and money supply policies are chosen such that the
expected joint utility obtained over a specified time horizon is
maximised, subject to a budget constraint that takes into account
the ``value" of the liquidity benefit associated with the money
supply. For any choice of the bivariate utility function, the
resulting model determines a relation between the rate of
consumption, the price level, and the money supply. The model
produces explicit expressions for the real and nominal pricing
kernels, and hence establishes a basis for the valuation of
inflation-linked securities.

\section{Asset pricing in discrete time}\label{sec7.1}
The development of asset-pricing theory in discrete time has been
pursued by many authors. In the context of interest rate
modelling, it is worth recalling that the first example of a fully
developed term-structure model where the initial discount function
is freely specifiable is that of Ho \& Lee 1986, in a
discrete-time setting. For our purposes it will be useful to
develop a general discrete-time scheme from first principles,
taking an axiomatic approach in the spirit of Hughston \&
Rafailidis (2005).

Let $\{t_i\}_{i=0,1,2,\ldots}$ denote a sequence of discrete
times, where $t_0$ represents the present and $t_{i+1}>t_i$ for
all $i\in\N_0$. We assume that the sequence $\{t_i\}$ is unbounded
in the sense that for any given time $T$ there exists a value of
$i$ such that $t_i>T$. We do not assume that the elements of
$\{t_i\}$ are equally spaced; for some applications, however, we
can consider the case where $t_n=n\tau$ for all $n\in\mathbb{N}_0$
and for some unit of time $\tau$.

Each asset is characterised by a pair of processes
$\{S_{t_i}\}_{i\ge 0}$ and $\{D_{t_i}\}_{i\ge 0}$ which we refer
to as the ``value process" and the ``dividend process",
respectively. We interpret $D_{t_i}$ as a random cash flow or
dividend paid to the owner of the asset at time $t_i$. Then
$S_{t_i}$ denotes the ``ex-dividend" value of the asset at $t_i$.
We can think of $S_{t_i}$ as the cash flow that would result if
the owner were to dispose of the asset at time $t_i$.

For simplicity, we shall frequently use an abbreviated notation,
and write $S_i=S_{t_i}$ and $D_i=D_{t_i}$. Thus $S_i$ denotes the
value of the asset at time $t_i$, and $D_i$ denotes the dividend
paid at time $t_i$. Both $S_i$ and $D_i$ are expressed in nominal
terms, i.e. in units of a fixed base currency. We use the term
``asset" in the broad sense here---the scheme is thus applicable
to any liquid financial position for which the values and cash
flows are well defined, and for which the principles of no
arbitrage are applicable.

The unfolding of random events in the economy will be represented
with the specification of a probability space $(\Omega,
\mathcal{F}, \mathbb{P})$ equipped with a filtration
$\{\mathcal{F}_i\}_{i\ge 0}$ which we call the ``market
filtration". For the moment we regard the market filtration as
given, but later we shall construct it explicitly. For each asset
we assume that the associated value and dividend processes are
adapted to $\{\F_i\}$. In what follows ${\mathbb{P}}$ is taken to
be the ``physical" or ``objective" probability measure; all
equalities and inequalities between random variables are to be
understood as holding almost surely with respect to $\mathbb{P}$.
For convenience we often write $\E_i[-]$ for $\E[-|\F_i]$.

In order to ensure the absence of arbitrage in the financial
markets and to establish intertemporal pricing relations, we
assume the existence of a strictly positive pricing kernel
$\{\pi_i\}_{i\ge 0}$, and make the following assumptions:
\vspace{.2cm}

\noindent {\bf Axiom~A}. {\it For any asset with associated value
process $\{S_i\}_{i\ge 0}$ and dividend process $\{D_i\}_{i\ge
0}$, the process $\{M_i\}_{i\ge 0}$ defined by
\begin{equation}\label{2.1}
M_i=\pi_i S_i+\sum^{i}_{n=0}\pi_n D_n
\end{equation}
is a martingale, i.e. $\E[|M_i|]<\infty$ for all
$i\in\mathbb{N}_0$, and $\E[M_j|\F_i]=M_i$ for all $i\le j$. }

\vspace{0.2cm} \noindent {\bf Axiom~B}. {\it There exists a
strictly positive non-dividend-paying asset, with value process
$\{\bar{B}_i\}_{i\ge 0}$, that offers a strictly positive return,
i.e.~such that $\bar{B}_{i+1}>\bar{B}_i$ for all
$i\in\mathbb{N}_0$. We assume that the process $\{\bar{B}_i\}$ is
unbounded in the sense that for any $b\in\mathbb{R}$ there exists
a time $t_i$ such that $\bar{B}_i>b$.} \vspace{.2cm}

Given this axiomatic scheme, we proceed to explore its
consequences. The notation $\{\bar{B}_i\}$ is used in Axiom B to
distinguish the positive return asset from the previsible
money-market account asset $\{B_i\}$ that will be introduced
later; in particular, in Proposition 6 it will be shown that
Axioms A and B imply the existence of a unique money-market
account asset.  We note that since the positive-return asset is
non-dividend paying, it follows from Axiom A that $\{\pi_i
\bar{B}_i\}$ is a martingale. Writing $\bar{\rho}_i=\pi_i
\bar{B}_i$, we have $\pi_i=\bar{\rho}_i/\bar{B}_i$. Since
$\{\bar{B}_i\}$ is assumed to be strictly increasing, we see that
$\{\pi_i\}$ is a supermartingale. In fact, we have the somewhat
stronger relation $\E_i[\pi_j]<\pi_i$. Indeed, we note that
\begin{equation}
\E_i[\pi_j]=\E_i\left[\frac{\bar{\rho}_j}{\bar{B}_j}\right]<E_i\left[\frac{\bar{\rho}_j}{\bar{B}_i}\right]=\frac{E_i[\bar{\rho}_j]}{\bar{B}_i}=\frac{\bar{\rho}_i}{\bar{B}_i}=\pi_i.
\end{equation}
The significance of $\{\bar{\rho}_i\}$ is that it has the
interpretation of being the likelihood ratio appropriate for a
change of measure from the objective measure $\mathbb{P}$ to the
equivalent martingale measure $\mathbb{Q}$ characterised by the
property that non-dividend-paying assets when expressed in units
of the numeraire $\{\bar{B}_i\}$ are martingales.

We recall the definition of a potential. An adapted process
$\{x_i\}_{0\le i<\infty}$ on a probability space $(\Omega, \F,
\mathbb{P})$ with filtration $\{\F_i\}$ is said to be a potential
if $\{x_i\}$ is a non-negative supermartingale and
$\lim_{i\rightarrow\infty}\E[x_i]=0$. It is straightforward to
show that $\{\pi_i\}$ is a potential. We need to demonstrate that
given any $\epsilon>0$ we can find a time $t_j$ such
$\E[\pi_n]<\epsilon$ for all $n\ge j$. This follows from the
assumption that the positive-return asset price process
$\{\bar{B}_i\}$ is unbounded in the sense specified in Axiom B.
Thus given $\epsilon$ let us set $b=\bar{\rho}_0/\epsilon$. Now
given $b$ we can find a time $t_j$ such that $\bar{B}_{t_n}>b$ for
all $n\ge j$. But for that value of $t_j$ we have
\begin{equation}
\E[\pi_j]=\E\left[\frac{\bar{\rho}_j}{\bar{B}_j}\right]<\frac{\E[\bar{\rho}_j]}{b}=\frac{\bar{\rho}_0}{b}=\epsilon,
\end{equation}
and hence $\E[\pi_n]<\epsilon$ for all $n\ge j$. It follows that
\begin{equation}\label{3.5aa}
\lim_{i\rightarrow\infty}\E[\pi_i]=0.
\end{equation}

Next we recall the Doob decomposition for discrete-time
supermartingales (see, e.g., Meyer 1966, chapter 7). If $\{x_i\}$
is a supermartingale on a probability space
$(\Omega,\F,\mathbb{P})$ with filtration $\{\F_i\}$, then there
exists a martingale $\{y_i\}$ and a previsible increasing process
$\{a_i\}$ such that $x_i=y_i-a_i$ for all $i\ge 0$. By previsible,
we mean that $a_i$ is $\F_{i-1}$-measurable. The decomposition is
given explicitly by $a_0=0$ and
$a_i=a_{i-1}+x_{i-1}-\E_{i-1}[x_i]$ for $i\ge 1$.

It follows that the pricing kernel admits a decomposition of this
form, and that one can write
\begin{equation}\label{DoobPI}
\pi_i=Y_i-A_i,
\end{equation}
where $A_0=0$ and
\begin{equation}\label{DoobDecomp}
A_i=\sum^{i-1}_{n=0}\left(\pi_n-\E_n[\pi_{n+1}]\right)
\end{equation}
for $i\ge 1$; and where $Y_0=\pi_0$ and
\begin{equation}
Y_i=\sum^{i-1}_{n=0}\left(\pi_{n+1}-\E_n[\pi_{n+1}]\right)+\pi_0
\end{equation}
for $i\ge 1$. The Doob decomposition for $\{\pi_i\}$ has an
interesting expression in terms of discount bonds, which we shall
mention later, in Section V.

In the case of a potential $\{x_i\}$ it can be shown (see, e.g.,
Gihman \& Skorohod 1979, chapter 1) that the limit
$a_{\infty}=\lim_{i\rightarrow\infty}a_i$ exists, and that
$x_i=\E_i[a_{\infty}]-a_i$. As a consequence, we conclude that the
pricing kernel admits a decomposition of the form
\begin{equation}
\pi_i=\E_i[A_{\infty}]-A_i,
\end{equation}
where $\{A_i\}$ is the previsible process defined by
(\ref{DoobDecomp}). With these facts in hand, we shall establish a
useful result concerning the pricing of limited-liability assets.
By a limited-liability asset we mean an asset such that $S_i\ge0$
and $D_i\ge0$ for all $i\in\mathbb{N}$.

\vspace{.4cm} \noindent{\bf Proposition 1}. {\it Let
$\{S_i\}_{i\ge 0}$ and $\{D_i\}_{i\ge 0}$ be the value and
dividend processes associated with a limited-liability asset. Then
$\{S_i\}$ is of the form
\begin{equation}\label{2.2}
S_i=\frac{m_i}{\pi_i}+\frac{1}{\pi_i}\E_i\left[\sum^{\infty}_{n=i+1}\pi_n
D_n\right],
\end{equation}
where $\{m_i\}$ is a non-negative martingale that vanishes if and
only if the following transversality condition holds:
\begin{equation}\label{2.3}
\lim_{j\rightarrow\infty}\E[\pi_j S_j]=0.
\end{equation}
}

\noindent{\bf Proof}. It follows from Axiom A, as a consequence of
the martingale property, that
\begin{equation}\label{2.4}
\pi_i S_i+\sum^i_{n=0}\pi_n D_n=\E_i\left[\pi_j
S_j+\sum^j_{n=0}\pi_n D_n\right]
\end{equation}
for all $i\le j$. Taking the limit $j\rightarrow\infty$ on the
right-hand side of this relation we have
\begin{equation}\label{2.5}
\pi_i S_i+\sum^i_{n=0}\pi_n
D_n=\lim_{j\rightarrow\infty}\E_i[\pi_j
S_j]+\lim_{j\rightarrow\infty}\E_i\left[\sum^j_{n=0}\pi_n
D_n\right].
\end{equation}
Since $\pi_i D_i\ge 0$ for all $i\in\mathbb{N}_0$, it follows from
the conditional form of the monotone convergence theorem---see,
e.g., Steele 2001, Williams 1991---that
\begin{equation}\label{2.6}
\lim_{j\rightarrow\infty}\E_i\left[\sum^j_{n=0}\pi_n
D_n\right]=\E_i\left[\lim_{j\rightarrow\infty}\sum^j_{n=0}\pi_n
D_n\right],
\end{equation}
and hence that
\begin{equation}\label{2.7}
\pi_i S_i+\sum^i_{n=0}\pi_n
D_n=\lim_{j\rightarrow\infty}\E_i[\pi_j
S_j]+\E_i\left[\sum^{\infty}_{n=0}\pi_n D_n\right].
\end{equation}
Now let us define
\begin{equation}\label{2.8}
m_i=\lim_{j\rightarrow\infty}\E_i[\pi_j S_j].
\end{equation}
Then clearly $m_i\ge 0$ for all $i\in\mathbb{N}_0$. We see,
moreover, that $\{m_i\}_{i\ge 0}$ is a martingale, since
$m_i=M_i-\E_i[F_{\infty}]$, where $M_i$ is defined as in equation
(\ref{2.1}), and
\begin{equation}\label{2.9}
F_{\infty}=\sum^{\infty}_{n=0}\pi_n D_n.
\end{equation}
It is implicit in the axiomatic scheme that the sum
$\sum^{\infty}_{n=0}\pi_n D_n$ converges in the case of a
limited-liability asset. This follows as a consequence of the
martingale convergence theorem and Axiom A. Thus, writing equation
(\ref{2.7}) in the form
\begin{equation}
\pi_i S_i+\sum^i_{n=0}\pi_n
D_n=m_i+\E_i\left[\sum^{\infty}_{n=0}\pi_n D_n\right],
\end{equation}
after some re-arrangement of terms we obtain
\begin{equation}\label{2.10}
\pi_i S_i=m_i+\E_i\left[\sum^{\infty}_{n=i+1}\pi_n D_n\right],
\end{equation}
and hence (\ref{2.2}), as required. On the other hand, by the
martingale property of $\{m_i\}$ we have $\E[m_i]=m_0$ and hence
\begin{equation}\label{2.11}
\E[m_i]=\lim_{j\rightarrow\infty}\E[\pi_j S_j]
\end{equation}
for all $i\in\N$. Thus since $m_i\ge 0$ we see that the
transversality
condition (\ref{2.3}) holds if and only if $\{m_i\}=0$.\hfill$\Box$\\

The interpretation of the transversality condition is as follows.
For each $j\in\N_0$ the expectation $V_j=\E[\pi_j S_j]$ measures
the present value of an instrument that pays at $t_j$ an amount
equal to the proceeds of a liquidation of the asset with price
process $\{S_i\}_{i\ge 0}$. If $\lim_{j\rightarrow\infty}V_j=0$
then one can say that in the long term all of the value of the
asset will be dispersed in its dividends. On the other hand, if
some or all of the dividends are ``retained" indefinitely, then
$\{V_j\}$ will retain some value, even in the limit as $t_j$ goes
to infinity.

The following example may clarify this interpretation. Suppose
investors put \$100m of capital into a new company. The management
of the company deposit \$10m into a money market account. The
remaining \$90m is invested in a risky line of business,
the proceeds of which, after costs, are paid to
share-holders as dividends. At time $t_i$ we have
$S_i=B_i+H_i$, where $B_i$ is a position in the money market
account initialised at \$10m, and $H_i$ is the value of the
remaining dividend flow. Now $\{\pi_i B_i\}$ is a martingale, and
thus $\E[\pi_i B_i]=\$10{\text m}$ for all $i\in\N_0$, and
hence $\lim_{i\rightarrow\infty}\E[\pi_i B_i]=\$10{\text m}$.
On the other hand $\lim_{i\rightarrow\infty}\E[\pi_i H_i]=0$; this
means that given any value $h$ we can find a time $T$ such that
$\E[\pi_i H_i]<h$ for all $t_i\ge T$.

There are other ways of ``retaining" funds than putting them into
a domestic money market account. For example, one could put the
\$10m into a foreign bank account; or one could invest it in
shares in a securities account, with a standing order that
dividends should be immediately re-invested in further shares.
If the investment is in a general ``dividend-retaining" asset
(such as a foreign bank account), then $\{m_i\}$ can in principal
be any non-negative martingale. The content of Proposition 1 is
that a limited-liability investment can be separated in a unique
way into a growth component and an income component.

In the case of a ``pure income" investment, i.e. in an asset for
which the transversality condition is satisfied, the price is
directly related to the future dividend flow, and we have
\begin{equation}\label{2.12}
S_i=\frac{1}{\pi_i}\E_i\left[\sum^{\infty}_{n=i+1}\pi_n
D_n\right].
\end{equation}
This is the so-called ``fundamental equation" which some authors
use directly as a basis for asset pricing theory---see, e.g.,
Cochrane 2005. Alternatively we can write (\ref{2.12}) in the form
\begin{equation}\label{2.13}
S_i=\frac{1}{\pi_i}(\E_i[F_{\infty}]-F_i),
\end{equation}
where
\begin{equation}\label{2.14}
F_i=\sum^i_{n=0}\pi_n D_n,\quad \textrm{and}\quad
F_{\infty}=\lim_{i\rightarrow\infty}F_i.
\end{equation}
It is straightforward to show that the process $\{\pi_i
S_i\}$ is a potential. Clearly, $\{\E_i[F_{\infty}]-F_i\}$ is a
positive supermartingale, since $\{F_i\}$ is increasing; and by
the tower property and the monotone convergence theorem we have
$\lim_{i\rightarrow\infty}\E[\E_i[F_{\infty}]-F_i]=\E[F_{\infty}]-\lim_{i\rightarrow\infty}\E[F_i]=\E[F_{\infty}]-\E\left[\lim_{i\rightarrow\infty}F_i\right]=0$.
On the other hand, $\{\pi_i\}$ is also a potential, so we conclude that in the case of an income generating asset the
price process can be expressed as a ratio of potentials, thus
giving us a discrete-time analogue of a result obtained by Rogers
1997. Indeed, the role of the concept of a potential as it appears
here is consistent with the continuous-time theories developed by
Flesaker \& Hughston 1996, Rogers 1997, Rutkowski 1997, Jin \&
Glasserman 2001, Hughston \& Rafailidis 2005, and others, where
similar structures arise.
\section{Nominal pricing kernel and nominal interest rates}\label{sec7.2}
\label{sec:3} To proceed further we need to say more about the
relation between the pricing kernel $\{\pi_i\}$ and the
positive-return asset $\{\bar{B}_i\}$. To this end let us write
\begin{equation}\label{3.1}
\bar{r}_i=\frac{\bar{B}_i-\bar{B}_{i-1}}{\bar{B}_{i-1}}
\end{equation}
for the rate of return on the positive-return asset realised at
time $t_i$ on an investment made at time $t_{i-1}$. Since the time
interval $t_i-t_{i-1}$ is not necessarily small, there is no
specific reason to presume that the rate of return $\bar{r}_i$ is
already known at time $t_{i-1}$. This is consistent with the fact
that we have assumed that $\{\bar{B}_i\}$ is $\{\F_i\}$-adapted.
The notation $\bar{r}_i$ is used here to distinguish the rate of
return on the positive-return asset from the rate of return $r_i$
on the money market account, which will be introduced in Section
V.

Next we present a simple argument to motivate the idea that there
should exist an asset with constant value unity that pays a
dividend stream given by $\{\bar{r}_i\}$. We consider the
following portfolio strategy. The portfolio consists at any time
of a certain number of units of the positive-return asset. Let
$\phi_i$ denote the number of units, so that at time $t_i$ the
(ex-dividend) value of the portfolio is given by
$V_i=\phi_i\bar{B}_i$. Then in order to have $V_i=1$ for all $i\ge
0$ we set $\phi_i=1/\bar{B}_i$. Let $D_i$ denote the dividend paid
out by the portfolio at time $t_i$. Then clearly if the portfolio
value is to remain constant we must have
$D_i=\phi_{i-1}\bar{B}_i-\phi_{i-1}\bar{B}_{i-1}$ for all $i\ge
1$. It follows immediately that $D_i=\bar{r}_i$, where $\bar{r}_i$
is given by (\ref{3.1}).

This shows that we can construct a portfolio with a constant value
and with the desired cash flows. Now we need to show that such a
system satisfies Axiom A.

\vspace{.4cm} \noindent{\bf Proposition 2}. {\it There exists an
asset with constant nominal value $S_i=1$ for all $i\in\N_0$, for
which the associated cash flows are given by $\{\bar{r}_i\}_{i\ge
1}$.}\\

\noindent{\bf Proof}. We need to verify that the conditions of
Axiom A are satisfied in the case for which $S_i=1$ and
$D_i=\bar{r}_i$ for $i\in\N_0$. In other words we need to show
that
\begin{equation}\label{3.2}
\pi_i=\E_i[\pi_j]+\E_i\left[\sum^j_{n=i+1}\pi_n \bar{r}_n\right]
\end{equation}
for all $i\le j$. The calculation proceeds as follows. We observe
that
\begin{eqnarray}\label{3.3}
\E_i\left[\sum^j_{n=i+1}\pi_n \bar{r}_n\right]&=&\E_i\left[\sum^j_{n=i+1}\pi_n\frac{\bar{B}_n-\bar{B}_{n-1}}{\bar{B}_{n-1}}\right]\nonumber\\
&=&\E_i\left[\sum^j_{n=i+1}\frac{\bar{\rho}_n}{\bar{B}_n}\frac{\bar{B}_n-\bar{B}_{n-1}}{\bar{B}_{n-1}}\right]\nonumber\\
&=&\E_i\left[\sum^j_{n=i+1}\left(\frac{\bar{\rho}_n}{\bar{B}_{n-1}}-\frac{\bar{\rho}_n}{\bar{B}_n}\right)\right]\nonumber\\
&=&\E_i\left[\sum^j_{n=i+1}\left(\E_{n-1}\left[\frac{\bar{\rho}_n}{\bar{B}_{n-1}}\right]-\frac{\bar{\rho}_n}{\bar{B}_n}\right)\right],
\end{eqnarray}
the last step being achieved by use of the tower property. It
follows then by use of the martingale property of $\{\rho_n\}$
that:
\begin{eqnarray}
\E_i\left[\sum^j_{n=i+1}\pi_n \bar{r}_n\right]&=&\E_i\left[\sum^j_{n=i+1}\left(\frac{1}{\bar{B}_{n-1}}\E_{n-1}[\bar{\rho}_n]-\frac{\bar{\rho}_n}{\bar{B}_n}\right)\right]\nonumber\\
&=&\E_i\left[\sum^j_{n=i+1}\left(\frac{\bar{\rho}_{n-1}}{\bar{B}_{n-1}}-\frac{\bar{\rho}_n}{\bar{B}_n}\right)\right]\nonumber\\
&=&\E_i\left[\left(\frac{\bar{\rho}_i}{\bar{B}_i}-\frac{\bar{\rho}_{i+1}}{\bar{B}_{i+1}}\right)+\left(\frac{\bar{\rho}_{i+1}}{\bar{B}_{i+1}}-\frac{\bar{\rho}_{i+2}}{\bar{B}_{i+2}}\right)
+\ldots+\left(\frac{\bar{\rho}_{j-1}}{\bar{B}_{j-1}}-\frac{\bar{\rho}_j}{\bar{B}_j}\right)\right]\nonumber\\
&=&\E_i\left[\frac{\bar{\rho}_i}{\bar{B}_i}\right]-\E_i\left[\frac{\bar{\rho}_j}{\bar{B}_j}\right]\nonumber\\
&=&\pi_i-\E_i[\pi_j].
\end{eqnarray}
But that gives us (\ref{3.2}). \hfill$\Box$\\

The existence of the constant-value asset leads to an alternative
decomposition of the pricing kernel, which can be described as
follows.

\vspace{.4cm} \noindent{\bf Proposition 3}. {\it Let
$\{\bar{B}_i\}$ be a positive-return asset satisfying the
conditions of Axiom B, and let $\{\bar{r}_i\}$ be its
rate-of-return process. Then the pricing kernel can be expressed
in the form $\pi_i=\E_i[G_{\infty}]-G_i$, where
$G_i=\sum^i_{n=1}\pi_n\bar{r}_n$ and
$G_{\infty}=\lim_{i\rightarrow\infty}G_i$.}\\

\noindent {\bf Proof}. First we remark that if an asset has
constant value then it satisfies the transversality condition
(\ref{2.3}). In particular, letting the constant be unity, we see
that the transversality condition reduces to
\begin{equation}\label{3.5a}
\lim_{i\rightarrow\infty}\E[\pi_i]=0,
\end{equation}
which is satisfied since $\{\pi_i\}$ is a potential. Next we show
that
\begin{equation}\label{3.5b}
\lim_{j\rightarrow\infty}\E_i[\pi_j]=0
\end{equation}
for all $i\in\N_0$. In particular, fixing $i$, we have
$\E\left[\E_i[\pi_j]\right]=\E[\pi_j]$ by the tower property, and
thus
\begin{equation}
\lim_{j\rightarrow\infty}\E\left[\E_i[\pi_j]\right]=0
\end{equation}
by virtue of (\ref{3.5a}). But $\E_i[\pi_j]<\pi_i$ for all $j>i$,
and $\E[\pi_i]<\infty$; hence by the dominated convergence theorem
we have
\begin{equation}
\lim_{j\rightarrow\infty}\E[\E_i[\pi_j]]=\E[\lim_{j\rightarrow\infty}\E_i[\pi_j]],
\end{equation}
from which the desired result (\ref{3.5b}) follows, since the
argument of the expectation is non-negative. As a consequence of
(\ref{3.5b}) it follows from (\ref{3.2}) that
\begin{equation}\label{3.6}
\pi_i=\lim_{j\rightarrow\infty}\E_i\left[\sum^j_{n=i+1}\pi_n
\bar{r}_n\right],
\end{equation}
and thus by the monotone convergence theorem we have
\begin{eqnarray}\label{3.7}
\pi_i&=&\E_i\left[\sum^{\infty}_{n=i+1}\pi_n \bar{r}_n\right]\nn\\
        &=&\E_i\left[\sum^{\infty}_{n=1}\pi_n \bar{r}_n\right]-\sum^{i}_{n=1}\pi_n
        \bar{r}_n\nn\\
        &=&\E_i\left[G_{\infty}\right]-G_i,
\end{eqnarray}
and that gives us the result of the proposition.\hfill$\Box$\\

We shall establish a converse to this result, which allows us to
construct a system satisfying Axioms A and B from any
strictly-increasing non-negative adapted process that converges,
providing a certain integrability condition holds.

\vspace{.4cm} \noindent{\bf Proposition 4}. {\it Let
$\{G_i\}_{i\ge 0}$ be a strictly increasing adapted process
satisfying $G_0=0$, and $\E[G_{\infty}]<\infty$, where
$G_{\infty}=\lim_{i\rightarrow\infty}G_i$. Let the processes
$\{\pi_i\}$, $\{\bar{r}_i\}$, and $\{\bar{B}_i\}$, be defined by
$\pi_i=\E_i[G_{\infty}]-G_{i}$ for $i\ge 0$;
$\bar{r}_i=(G_i-G_{i-1})/\pi_i$ for $i\ge 1$;
$\bar{B}_i=\prod^i_{n=1}(1+\bar{r}_n)$ for $i\ge 1$, with
$\bar{B}_0=1$. Let the process $\{\bar{\rho}_i\}$ be defined by
$\bar{\rho}_i=\pi_i \bar{B}_i$ for $i\ge0$. Then
$\{\bar{\rho}_i\}$ is a martingale, and
$\lim_{j\rightarrow\infty}\bar{B}_j=\infty$. Thus $\{\pi_i\}$ and
$\{\bar{B}_i\}$, as constructed, satisfy Axioms A and B.}\\

\noindent{\bf Proof}. Writing $g_i=G_i-G_{i-1}$ for $i\ge1$ we
have
\begin{equation}\label{3.8}
\pi_i=\E_i[G_{\infty}]-G_i=\E_i\left[\sum^{\infty}_{n=i+1}g_n\right],
\end{equation}
and
\begin{eqnarray}\label{3.9}
\bar{B}_i=\prod^i_{n=1}(1+\bar{r}_n)=\prod^i_{n=1}\left(1+\frac{g_n}{\pi_n}\right)=\prod^i_{n=1}\left(\frac{\pi_n+g_n}{\pi_n}\right).
\end{eqnarray}
Hence, writing $\bar{\rho}_i=\pi_i \bar{B}_i$, we have
\begin{eqnarray}\label{3.10}
\bar{\rho}_i&=&\pi_i\prod^i_{n=1}\left(\frac{\pi_n+g_n}{\pi_n}\right)\nn\\
&=&(\pi_i+g_i)\prod^{i-1}_{n=1}\left(\frac{\pi_n+g_n}{\pi_n}\right),
\end{eqnarray}
and thus
\begin{equation}
\bar{\rho}_i=(\pi_i+g_i)\bar{B}_{i-1}=\frac{\pi_i+g_i}{\pi_{i-1}}\bar{\rho}_{i-1}.
\end{equation}
To show that $\{\bar{\rho}_i\}$ is a martingale it suffices to
verify for all $i\ge 1$ that $\E[\bar{\rho}_i]<\infty$ and that
$\E_{i-1}[\bar{\rho}_i]=\bar{\rho}_{i-1}$. In particular, if
$\E[\bar{\rho}_i]<\infty$ then the ``take out what is known rule''
applies, and by (\ref{3.8}) and (\ref{3.10}) we have
\begin{eqnarray}
\E_{i-1}[\bar{\rho}_i]&=&\E_{i-1}\left[\frac{\pi_i+g_i}{\pi_{i-1}}\bar{\rho}_{i-1}\right]\nn \\
             &=&\frac{\bar{\rho}_{i-1}}{\pi_{i-1}}\E_{i-1}\left[\pi_i+g_i\right]\nn\\
             &=&\frac{\bar{\rho}_{i-1}}{\pi_{i-1}}\E_{i-1}\left[\sum^{\infty}_{n=i}g_n\right]\nn\\
             &=&\frac{\bar{\rho}_{i-1}}{\pi_{i-1}}\left(\E_{i-1}\left[G_{\infty}\right]-G_{i-1}\right)\nn\\
             &=&\bar{\rho}_{i-1}.
\end{eqnarray}
Here, in going from the first to the second line we have used the
fact that $\E[\pi_i+g_i]<\infty$, together with the assumption
that $\E[\bar{\rho}_i]<\infty$. To verify that
$\E[\bar{\rho}_i]<\infty$ let us write
\begin{equation}
J^{\alpha}_{i-1}=\min\left[\frac{\bar{\rho}_{i-1}}{\pi_{i-1}},\alpha\right]
\end{equation}
for $\alpha\in\mathbb{N}_0$. Then by use of monotone convergence
and the tower property we have
\begin{eqnarray}
\E[\bar{\rho}_i]&=&\E\left[(\pi_i+g_i)\lim_{\alpha\rightarrow\infty}J^{\alpha}_{i-1}\right]\nn\\
        &=&\lim_{\alpha\rightarrow\infty}\E\left[(\pi_i+g_i)J^{\alpha}_{i-1}\right]\nn\\
        &=&\lim_{\alpha\rightarrow\infty}\E\left[\E_{i-1}\left[(\pi_i+g_i)J^{\alpha}_{i-1}\right]\right]\nn\\
        &=&\lim_{\alpha\rightarrow\infty}\E\left[J^{\alpha}_{i-1}\E_{i-1}\left[(\pi_i+g_i)\right]\right]\nn\\
        &\le&\E\left[\frac{\bar{\rho}_{i-1}}{\pi_{i-1}}\E_{i-1}[\pi_i+g_i]\right]\nn\\
        &=&\E[\bar{\rho}_{i-1}],
\end{eqnarray}
since
\begin{equation}
J^{\alpha}_{i-1}\le\frac{\bar{\rho}_{i-1}}{\pi_{i-1}}.
\end{equation}
Thus we see for all $i\ge 1$ that if
$\E\left[\bar{\rho}_{i-1}\right]<\infty$ then
$\E\left[\bar{\rho}_i\right]<\infty$. But $\bar{\rho}_0<\infty$ by
construction; hence by induction we deduce that
$\E\left[\bar{\rho}_i\right]<\infty$ for all $i\ge 0$.

 To show that $\lim_{j\rightarrow\infty}\{\bar{B}_j\}=\infty$ let us assume the contrary and show that this
 leads to a contradiction. Suppose, in particular, that there were to exist a number $b$ such that
 $\bar{B}_i<b$ for all $i\in\N_0$. Then for all $i\in\N_0$ we would have
\begin{equation}\label{3.13a}
\E\left[\frac{\bar{\rho}_i}{\bar{B}_i}\right]>\frac{1}{b}\E[\bar{\rho}_i]=\frac{\bar{\rho}_0}{b}.
\end{equation}
But by construction we know that
$\lim_{i\rightarrow\infty}\E[\pi_i]=0$ and hence
\begin{equation}
\lim_{i\rightarrow\infty}\E\left[\frac{\bar{\rho}_i}{\bar{B}_i}\right]=0.
\end{equation}
Thus given any $\epsilon>0$ we can find a time $t_i$ such that
\begin{equation}\label{3.14}
\E\left[\frac{\bar{\rho}_i}{\bar{B}_i}\right]<\epsilon.
\end{equation}
But this is inconsistent with (\ref{3.13a}); and thus we conclude
that $\lim_{j\rightarrow\infty}\bar{B}_j=\infty$.
That completes the proof of Proposition 4.\hfill$\Box$\\
 \section{Nominal discount bonds}\label{sec7.3}
Now we proceed to consider the properties of nominal discount
bonds. By such an instrument we mean an asset that pays a single
dividend consisting of one  unit of domestic currency at some
designated time $t_j$. For the price $P_{ij}$ at time $t_i$
$(i<j)$ of a discount bond that matures at time $t_j$ we thus have
\begin{equation}\label{3.15}
P_{ij}=\frac{1}{\pi_i}\E_i[\pi_j].
\end{equation}
Since $\pi_i>0$ for all $i\in\N$, and $\E_i[\pi_j]<\pi_i$ for all
$i<j$, it follows that $0<P_{ij}<1$ for all $i<j$. We observe, in
particular, that the associated interest rate $R_{ij}$ defined by
\begin{equation}\label{3.16}
P_{ij}=\frac{1}{1+R_{ij}}
\end{equation}
is strictly positive. In our theory we regard a discount
bond as a ``dividend-paying" asset. Thus in the case of a discount
bond with maturity $t_j$ we have $P_{jj}=0$ and $D_j=1$. Usually
discount bonds are defined by setting $P_{jj}=1$ at maturity, with
$D_j=0$; but it is more logical to regard the bonds as
giving rise to a unit cash flow at maturity. The
definition of the discount bond system does not involve the
specific choice of the positive-return asset.

It is important to point out that in the present framework there
is no reason or need to model the dynamics of $\{P_{ij}\}$, or to
model the volatility structure of the discount bonds. Indeed, from
the present point of view this would be a little artificial. The
important issue, rather, is how to model the pricing kernel. Thus,
our scheme differs somewhat in spirit from the discrete-time
models discussed, e.g., in Heath {\it et al.} 1990, and
Filipovi\'c \& Zabczyk 2002.

As a simple example of a family of discrete-time interest rate
models admitting tractable formulae for the associated discount
bond price processes, suppose we set
\begin{equation}\label{3.17}
\pi_i=\alpha_i+\beta_i N_i
\end{equation}
where $\{\alpha_i\}$ and $\{\beta_i\}$ are strictly-positive,
strictly-decreasing deterministic sequences, satisfying
$\lim_{i\rightarrow\infty}\alpha_i=0$ and
$\lim_{i\rightarrow\infty}\beta_i=0$, and where $\{N_i\}$ is a
strictly positive martingale. Then by (\ref{3.15}) we have
\begin{eqnarray}\label{3.18}
P_{ij}=\frac{\alpha_j+\beta_j N_i}{\alpha_i+\beta_i N_i},
\end{eqnarray}
thus giving a family of ``rational" interest rate models. Note
that in a discrete-time setting we can produce classes of models
that have no immediate analogues in continuous time---for example,
we can let $\{N_i\}$ be the natural martingale associated with a
branching process.

Now we shall demonstrate that any discount bond system consistent
with our general scheme admits a representation of the
Flesaker-Hughston type. For accounts of the Flesaker-Hughston
theory see, e.g., Flesaker \& Hughston 1996, Rutkowski 1997, Hunt
\& Kennedy 2000, or Jin \& Glasserman 2001.

\vspace{.4cm} \noindent{\bf Proposition 5}. {\it Let $\{\pi_i\}$,
$\{\bar{B}_i\}$, $\{P_{ij}\}$ satisfy the conditions of {\rm
Axioms A} and {\rm B}. Then there exists a family of positive
martingales $\{m_{in}\}_{0\le i\le n}$ indexed by $n\in\N$ such
that
\begin{equation}\label{3.19}
P_{ij}=\frac{\sum^{\infty}_{n=j+1}m_{in}}{\sum^{\infty}_{n=i+1}m_{in}}.
\end{equation}
}

\noindent{\bf Proof}. We shall use the fact that $\pi_i$ can be
written in the form
\begin{eqnarray}\label{3.20}
\pi_i&=&\E_i[G_{\infty}]-G_{i}\nonumber\\
       &=&\E_i\left[\sum^{\infty}_{n=1}g_n\right]-\sum^i_{n=1}g_n\nonumber\\
       &=&\E_i\left[\sum^{\infty}_{n=i+1}g_n\right],
\end{eqnarray}
where $g_i=G_i-G_{i-1}$ for each $i\ge 1$. Then $g_i>0$ for all
$i\ge 1$ since $\{G_i\}$ is a strictly increasing sequence. By the
monotone convergence theorem we have
\begin{equation}\label{3.21}
\pi_i=\sum^{\infty}_{n=i+1}\E_i[g_n]
\end{equation}
and
\begin{equation}\label{3.22}
\E_i[\pi_j]=\sum^{\infty}_{n=j+1}\E_i[g_n].
\end{equation}
For each $n\ge 1$ we define $m_{in}=\E_i[X_n]$. Then for each
$n\in\N$ we see that $\{m_{in}\}_{0\le i\le n}$ is a strictly
positive martingale, and (\ref{3.19}) follows
immediately.\hfill$\Box$
\section{Nominal money-market account}\label{sec7.4}
In the analysis presented so far we have assumed that the
positive-return process $\{\bar{B}_i\}$ is $\{F_i\}$-adapted, but
is not necessarily previsible. Many of our
conclusions are valid under the weaker hypothesis of mere
adaptedness, as we have seen. There are also economic motivations
behind the use of the more general assumption. One can imagine
that the time sequence $\{t_i\}$ is in reality a ``course
graining" of a finer time sequence that includes the original
sequence as a sub-sequence. Then likewise one can imagine that
$\{\bar{B}_i\}$ is a sub-sequence of a finer process that assigns
a value to the positive-return asset at each time in the finer
time sequence. Finally, we can imagine that $\{\F_i\}$ is a
sub-filtration of a finer filtration based on the finer sequence.
In the case of a money market account, where the rate of interest
is set at the beginning of each short deposit period, we would like to regard the relevant value process as being
previsible with respect to the finer filtration, but merely
adapted with respect to the course-grained filtration.

Do positive-return assets, other than the standard previsible
money market account, actually exist in a discrete-time setting?
The following example gives an affirmative answer. In the setting
of the standard binomial model, in the case of a single period,
let $S_0$ denote the value at time $0$ of a risky asset, and let
$\{U,D\}$ denote its possible values at time $1$. Let $B_0$ and
$B_1$ denote the values at times 0 and 1 of a deterministic
money-market account. We assume that $B_1>B_0$ and $U>S_0
B_1/B_0>D$. A standard calculation shows that the risk-neutral
probabilities for $S_0\rightarrow U$ and $S_0\rightarrow D$ are
given by $p^*$ and $1-p^*$, where $p^*=(S_0 B_1/B_0-D)/(U-D)$. We
shall now construct a ``positive-return" asset, i.e. an asset with
initial value $\bar{S}_0$ and with possible values
$\{\bar{U},\bar{D}\}$ at time 1 such that $\bar{U}>\bar{S}_0$ and
$\bar{D}>\bar{S}_0$. Risk-neutral valuation implies that
$\bar{S}_0=(B_0/B_1)[p^*\bar{U}+(1-p^*)\bar{D}]$. Thus, given
$\bar{S}_0$, we can determine $\bar{U}$ in terms of $\bar{D}$. A
calculation then shows that if
$(B_1/B_0-p^*)/(1-p^*)>\bar{D}/\bar{S}_0>1$, then
$\bar{U}>\bar{S}_0$ and $\bar{D}>\bar{S}_0$, as desired. Thus, in
the one-period binomial model, for the given initial value
$\bar{S}_0$, we obtain a one-parameter family of positive-return
assets.

Let us consider now the special case where the positive-return
asset is previsible. Thus for $i\ge 1$ we assume that $B_i$ is
$\F_{i-1}$-measurable and we drop the ``bar" over $B_i$ to signify
the fact that we are now considering a money-market account. In
that case we have
\begin{eqnarray}\label{3.23}
P_{i-1,i}&=&\frac{1}{\pi_{i-1}}\E_{i-1}[\pi_i]\nonumber\\
        &=&\frac{B_{i-1}}{\rho_{i-1}}\E_{i-1}\left[\frac{\rho_i}{B_i}\right]\nonumber\\
        &=&\frac{B_{i-1}}{B_i},
\end{eqnarray}
by virtue of the martingale property of $\{\rho_i\}$. Thus, in the
case of a money-market account we see that
\begin{equation}\label{3.24}
P_{i-1,i}=\frac{1}{1+r_i}.
\end{equation}
where $r_i=R_{i-1,i}$. In other words, the rate of return on the
money-market account is previsible, and is given by the one-period
simple discount factor associated with the discount bond that
matures at time $t_i$.

Reverting now to the general situation, it follows that if we are
given a pricing kernel $\{\pi_i\}$ on a probability space
$(\Omega,\F,\mathbb{P})$ with filtration $\{\F_i\}$, and a system
of assets satisfying Axioms A and B, then we can construct a
plausible candidate for an associated previsible money market
account by setting $B_0=1$ and defining
\begin{equation}\label{3.25}
B_i=(1+r_i)(1+r_{i-1})\cdots(1+r_1),
\end{equation}
for $i\ge 1$, where
\begin{equation}
r_i=\frac{\pi_{i-1}}{\E_{i-1}[\pi_i]}-1.
\end{equation}
We shall refer to the process $\{B_i\}$ thus constructed as the
``natural" money market account associated with the pricing kernel
$\{\pi_i\}$.

To justify this nomenclature, we need to verify that $\{B_i\}$, so
constructed, satisfies the conditions of Axioms A and B. To this
end, we make note of the following decomposition. Let $\{\pi_i\}$
be a positive supermartingale satisfying $\E_i[\pi_j]<\pi_i$ for
all $i<j$ and $\lim_{j\rightarrow\infty}[\pi_j]=0$. Then as an
identity we can write
\begin{equation}\label{3.26}
\pi_i=\frac{\rho_i}{B_i},
\end{equation}
where
\begin{equation}\label{3.27}
\rho_i=\frac{\pi_i}{\E_{i-1}[\pi_i]}\
\frac{\pi_{i-1}}{\E_{i-2}[\pi_{i-1]}}\cdots\frac{\pi_1}{\E_0[\pi_1]}\pi_0
\end{equation}
for $i\ge 0$, and
\begin{equation}\label{3.28}
B_i=\frac{\pi_{i-1}}{\E_{i-1}[\pi_i]}\
\frac{\pi_{i-2}}{\E_{i-2}[\pi_{i-1}]}\cdots\frac{\pi_1}{\E_1[\pi_2]}\
\frac{\pi_0}{\E_0[\pi_1]}
\end{equation}
for $i\ge 1$, with $B_0=1$. Thus, in this scheme we have
\begin{equation}\label{3.29}
\rho_i=\frac{\pi_i}{\E_{i-1}[\pi_i]}\rho_{i-1},
\end{equation}
with the initial condition $\rho_0=\pi_0$; and
\begin{equation}\label{3.30}
B_i=\frac{\pi_{i-1}}{\E_{i-1}[\pi_i]}B_{i-1},
\end{equation}
with the initial condition $B_0=1$. It is evident that
$\{\rho_i\}$ as thus defined is $\{\F_i\}$-adapted, and that
$\{B_i\}$ is previsible and strictly increasing. Making use of the
identity (\ref{3.30}) we are now in a position to establish the
following:

\vspace{.4cm} \noindent{\bf Proposition 6}. {\it Let $\{\pi_i\}$
be a non-negative supermartingale satisfying $\E_i[\pi_j]<\pi_i$
for all $i<j\in\N_0$, and $\lim_{i\rightarrow\infty}\E[\pi_i]=0$.
Let $\{B_i\}$ be defined by $B_0=1$ and $B_i=\prod^i_{n=1}(1+r_n)$
for $i\ge 1$, where $1+r_i=\pi_{i-1}/\E_{i-1}[\pi_i]$, and set
$\rho_i=\pi_i B_i$ for $i\ge 0$. Then $\{\rho_i\}$ is a
martingale, and the interest rate system defined by $\{\pi_i\}$,
$\{B_i\}$, and $\{P_{ij}\}$ satisfies {\rm Axioms A} and {\rm B}.
}\\

\noindent{\bf Proof}. To show that $\{\rho_i\}$ is a martingale it
suffices to verify for all $i\ge 1$ that $\E[\rho_i]<\infty$ and
that $\E_{i-1}[\rho_i]=\rho_{i-1}$. In particular, if
$\E[\rho_i]<\infty$ then the ``take out what is known rule" is
applicable, and by (\ref{3.29}) we have
\begin{equation}
\E_{i-1}[\rho_i]=\E_{i-1}\left[\frac{\pi_i}{\E_{i-1}[\pi_i]}\rho_{i-1}\right]=\rho_{i-1}.
\end{equation}
Thus to show that $\{\rho_i\}$ is a martingale all that remains is
to verify that $\E[\rho_i]<\infty$. Let us write
\begin{equation}
J^{\alpha}_{i-1}=\min\left[\frac{\rho_{i-1}}{\E_{i-1}[\pi_i]},\alpha\right]
\end{equation}
for $\alpha\in\N_0$. Then by monotone convergence and the tower
property we have
\begin{eqnarray}
\E[\rho_i]&=&\E\left[\pi_i\lim_{\alpha\rightarrow\infty}J^{\alpha}_{i-1}\right]\\
               &=&\lim_{\alpha\rightarrow\infty}\E\left[\pi_i J^{\alpha}_{i-1}\right]\\
               &=&\lim_{\alpha\rightarrow\infty}\E\left[\E_{i-1}[\pi_i J^{\alpha}_{i-1}]\right].
\end{eqnarray}
But since $J^{\alpha}_{i-1}$ is bounded we can move this term
outside the inner conditional expectation to give
\begin{equation}
\E[\rho_i]=\lim_{\alpha\rightarrow\infty}\E\left[J^{\alpha}_{i-1}\E_{1-i}[\pi_i]\right]\le\E[\rho_{i-1}],
\end{equation}
since
\begin{equation}
J^{\alpha}_{i-1}\le\frac{\rho_{i-1}}{\E_{i-1}[\pi_i]}.
\end{equation}
Thus we see for all $i\ge 1$ that if $\E[\rho_{i-1}]<\infty$ then
$\E[\rho_i]<\infty$. But $\rho_0<\infty$ by construction,
and hence by induction we deduce that $\E[\rho_i]<\infty$ for all $i\ge 0$.\hfill$\Box$\\

The martingale $\{\rho_i\}$ is the likelihood ratio process
appropriate for a change of measure from the objective measure
$\mathbb{P}$ to the equivalent martingale measure $\mathbb{Q}$
characterised by the property that non-dividend-paying assets are
martingales when expressed in units of the money-market account.
An interesting feature of Proposition 6 is that no integrability
condition is required on $\{\rho_i\}$. In other words, the natural
previsible money market account defined by (\ref{3.28})
``automatically" satisfies the conditions of Axiom A. For some
purposes it may therefore be advantageous to incorporate the
existence of the natural money market account directly into the
axioms. Then instead of Axiom B we would have:

\vspace{0.4cm} \noindent {\bf Axiom~B$^{\ast}$}. {{\it There
exists a strictly-positive non-dividend paying asset, the
money-market account, with value process $\{B_i\}_{i\ge 0}$,
having the properties that $B_{i+1}>B_i$ for all $i\in\N_0$ and
that $B_i$ is $\F_{i-1}$-measurable for all $i\in\N$. We assume
that $\{B_i\}$ is unbounded in the sense that for any
$b\in\mathbb{R}$ there exists a time $t_i$ such that $B_i>b$.}}
\vspace{0.2cm}

The content of Proposition 6 is that Axioms A and B together imply
Axiom B$^{\ast}$. As an exercise we shall establish that the class
of interest rate models satisfying Axioms ${\text A}$ and ${\text
B^{\ast}}$ is non-vacuus. In particular, suppose we consider the
``rational" models defined by equations (\ref{3.17}) and
(\ref{3.18}) for some choice of the martingale $\{N_i\}$. It is
straightforward to see that the unique previsible money market
account in this model is given by $B_0=1$ and
\begin{equation}
B_i=\prod^i_{n=1}\frac{\alpha_{n-1}+\beta_{n-1}
N_{n-1}}{\alpha_n+\beta_n N_{n-1}}
\end{equation}
for $i\ge 1$. For $\{\rho_i\}$ we then have
\begin{equation}
\rho_i=\rho_0\prod^{i}_{n=1}\frac{\alpha_n+\beta_n
N_n}{\alpha_n+\beta_n N_{n-1}},
\end{equation}
where $\rho_0=\alpha_0+\beta_0 N_0$. But it is easy to check that
for each $i\ge 0$ the random variable $\rho_i$ is bounded;
therefore $\{\rho_i\}$ is a martingale, and the money market
account process $\{B_i\}$ satisfies the conditions of Axioms A and
B$^{\ast}$.

Now let us return to the Doob decomposition for $\{\pi_i\}$ given
in formula (\ref{DoobPI}). Evidently, we have
$\pi_i=\E_i[A_{\infty}]-A_i$, with
\begin{eqnarray}
A_i&=&\sum^{i-1}_{n=0}\left(\pi_n-\E_n[\pi_{n+1}]\right)\nn\\
    &=&\sum^{i-1}_{n=0}\pi_n\left(1-\frac{\E_n[\pi_{n+1}]}{\pi_n}\right)\nn\\
    &=&\sum^{i-1}_{n=0}\pi_n\left(1-P_{n,n+1}\right)\nn\\
    &=&\sum^{i-1}_{n=0}\pi_n r_{n+1}P_{n,n+1},
\end{eqnarray}
where $\{r_i\}$ is the previsible short rate process defined by
(\ref{3.24}). The pricing kernel can therefore be put in the form
\begin{equation}\label{modPK}
\pi_i=\E_i\left[\sum^{\infty}_{n=i}\pi_n r_{n+1}P_{n,n+1}\right].
\end{equation}
Comparing the Doob decomposition (\ref{modPK}) with the
alternative decomposition given by (\ref{3.7}), we thus deduce
that if we set
\begin{equation}
\bar{r}_i=\frac{r_i\pi_{i-1}P_{i-1,i}}{\pi_i}
\end{equation}
then we obtain a positive-return asset for which the corresponding
decomposition of the pricing kernel, as given by (\ref{3.7}), is
the Doob decomposition. On the other hand, since the money-market
account is a positive-return asset, by Proposition 3 we can also
write
\begin{equation}
\pi_i=\E_i\left[\sum^{\infty}_{n=i+1}\pi_n r_n\right].
\end{equation}
As a consequence, we see that the price process of a pure income
asset can be written in the symmetrical form
\begin{equation}
S_i=\frac{\E_i\left[\sum^{\infty}_{n=i+1}\pi_n
D_n\right]}{\E_i\left[\sum^{\infty}_{n=i+1}\pi_n r_n\right]},
\end{equation}
where $\{D_n\}$ is the dividend process, and $\{r_n\}$ is the
short rate process.
\section{Information-based interest rate models}\label{sec7.5}
So far in the discussion we have regarded the pricing kernel
$\{\pi_i\}$ and the filtration $\{\F_i\}$ as being exogenously specified. To develop the framework further
we need to make a more specific indication of how the pricing
kernel is determined, and how information is made available
to market participants. To obtain a realistic model for
$\{\pi_i\}$ we need to develop the model in conjunction with a
theory of consumption, money supply, price level, inflation, real
interest rates, and information. We shall proceed in two steps.
First we consider a general ``reduced-form" model for nominal
interest rates, in which we model the filtration explicitly; then
in the next section we consider a more general ``structural''
model in which both the nominal and the real interest rate systems
are determined.

Our reduced-form model for interest rates is based on the
theory of $X$-factors, following Brody {\it et
al.} 2006 and Macrina 2006. Associated with each $t_i$ we
introduce a collection of random variables
$X^{\alpha}_i$ $(\alpha=1,\ldots, m_i)$, where $m_i$ denotes the
number of random variables associated with $t_i$. For each
$n$, we assume that the random variables
$X^{\alpha}_1, X^{\alpha}_2,\ldots,X^{\alpha}_n$ are independent.
We regard $X^{\alpha}_n$ as being ``revealed"
at time $t_n$, and hence $\F_n$-measurable. More precisely, we
shall construct the filtration $\{\F_i\}$ in such a way that this
property holds. Intuitively, we can think of $X^{\alpha}_1,
X^{\alpha}_2,\ldots,X^{\alpha}_n$ as being the independent
macroeconomic ``market factors" that determine cash flows at
$t_n$.

Let us consider how the filtration will be modelled. For each
$j\in\N_0$, at any time $t_i$ before $t_j$ only partial
information about the market factors $X^{\alpha}_j$ will be
available to market participants. We model this partial
information for each market factor $X^{\alpha}_j$ by defining a
discrete-time information process $\{\xi^{\alpha}_{t_i
t_j}\}_{0\le t_i\le t_j}$, setting
\begin{equation}
\xi^{\alpha}_{t_i t_j}=\sigma t_i
X^{\alpha}_{j}+\beta^{\alpha}_{t_i t_j}.
\end{equation}
Here $\{\beta^{\alpha}_{t_i t_j}\}_{0\le t_i\le t_j}$ can, for
each value of $\alpha$, be thought of as an independent
discretised Brownian bridge. Thus, we consider a standard Brownian
motion starting at time zero and ending at time $t_j$, and sample
its values at the times $\{t_i\}_{i=0,\ldots,j}$. Let us write
$\xi^{\alpha}_{ij}=\xi^{\alpha}_{t_i t_j}$ and
$\beta^{\alpha}_{ij}=\beta^{\alpha}_{t_i t_j}$, in keeping with
our usual shorthand conventions for discrete-time modelling.
For each value of $\alpha$ we have $\E[\beta^{\alpha}_{ij}]=0$ and
\begin{equation}
\textrm{Cov}[\beta^{\alpha}_{ik},\beta^{\alpha}_{jk}]=\frac{t_i(t_k-t_{j})}{t_k}
\end{equation}
for $i\le j\le k$. We assume that the bridge processes are
independent of the $X$-factors (i.e., the macroeconomic factors);
and hence that the information processes are independent
of one another. Finally, we assume that the market filtration is
generated collectively by the information processes. For
each value of $k$ the sigma-algebra $\F_k$ is generated by the
random variables $\{\xi^{\alpha}_{ij}\}_{0\le i\le j\le k}$.

Thus, the filtration is not simply ``given", but rather is
modelled explicitly. It is straightforward to verify
that for each value of $\alpha$ the process
$\{\xi^{\alpha}_{ij}\}$ has the Markov property. The proof follows
the pattern of the continuous-time argument. This has the
implication that the conditional expectation of a function of the
market factors $X^{\alpha}_j$, taken with respect to $\F_i$, can
be reduced to a conditional expectation with respect to the
sigma-algebra $\sigma(\xi^{\alpha}_{ij})$. Thus, the
history of the process $\{\xi^{\alpha}_{nj}\}_{n=0,1,\ldots,i}$
can be neglected, and only the most ``recent" information,
$\xi^{\alpha}_{ij}$, needs to be considered in taking the
conditional expectation.

For example, in the case of a function of a single
$\F_j$-measurable market factor $X_j$, with the associated
information process $\{\xi_{nj}\}_{n=0,1,\ldots,j}$, we obtain:
\begin{equation}
\E[f(X_j)|\F_i]=\frac{\int^{\infty}_0
p(x)f(x)\exp\left[\frac{t_j}{t_j-t_i}\left(\sigma
x\xi_{ij}-\frac{1}{2}\sigma^2 x^2 t_i\right)\right]\rd x}
{\int^{\infty}_0 p(x)\exp\left[\frac{t_j}{t_j-t_i}\left(\sigma
x\xi_{ij}-\frac{1}{2}\sigma^2 x^2 t_i\right)\right]\rd x},
\end{equation}
for $i\le j$, where $p(x)$ denotes the {\it a priori} probability
density function for the random variable $X_j$.

In the formula above we have presented the result
in the case of a single $X$-factor represented by a continuous
random variable taking non-negative values; the extension to other
classes of random variables is straightforward.

Now we are in a position to state how we propose to model the
pricing kernel. First, we shall assume that $\{\pi_i\}$ is adapted
to the market filtration $\{\F_i\}$. This is clearly a natural
assumption from an economic point of view, and is necessary for
the general consistency of the theory. This means that the random
variable $\pi_j$, for any fixed value of $j$, can be expressed as
a function of the totality of the available market information at
time $j$. In other words, $\pi_j$ is a function of the values
taken, between times $0$ and $j$, of the information processes
associated with the various market factors.

Next we make the assumption that $\pi_j$ (for any
fixed $j$) depends on the values of only a finite number of
information processes. This corresponds to the intuitive idea that
when we price a contingent claim, there is a limit to the
amount of information we can consider.

But this implies that expectations of the form $\E_i[\pi_j]$, for
$i\le j$, can be computed explicitly. Since
$\pi_j$ can be expressed as a function of a collection of
intertemporal information variables, the relevant conditional
expectations can be worked out in closed form. As a consequence, we are
led to a system of tractable expressions for the
discount bond prices and the previsible money market
account. We are left only with the question of what is the
correct functional form for $\{\pi_i\}$, given the relevant market
factors. If we simply ``postulate" a form for
$\{\pi_i\}$, then we say that we have a ``reduced-form" model. If
we provide an economic argument that leads to a specific form for
$\{\pi_i\}$, then we say that we have a ``structural" model.
\section{Models for inflation and index-linked securities}\label{sec7.6}
For a more complete picture we must regard the nominal interest
rate system as embedded in a larger system that takes into account
the macroeconomic factors that inter-relate the money
supply, aggregate consumption, and the price level. We shall
present a simple model in this spirit that is consistent with the
information-based approach.

To this end we introduce the following quantities. We envisage a
closed economy with aggregate consumption $\{k_i\}_{i\ge 1}$.
Consumption takes place at discrete times, and $k_i$ denotes the
aggregate level of consumption, in units of goods and services,
taking place at time $t_i$. We write $\{M_i\}_{i\ge 0}$ for
the process corresponding to the nominal money supply, and
$\{C_i\}_{i\ge 0}$ for the process of the consumer price index
(the ``price level"). For convenience we can regard $\{k_i\}$
and $\{M_i\}$ as being expressed on a {\it per capita} basis.
Hence these quantities can be regarded, respectively, as the
consumption and money balance associated with a representative
agent. We can thus formulate the optimisation problem from
the perspective of a representative agent; the role of the
agent here is to characterise the structure of the economy as a
whole.

We assume that at each $t_i$ the agent receives a
benefit or service from the money balance maintained in the
economy; this is given in nominal terms by $\lambda_i M_i$,
where $\lambda_i$ is the nominal liquidity benefit received
the agent per unit of money ``carried" by the agent, and $M_i$ is
the money supply, expressed on a {\it per capita} basis, at that
time. The corresponding ``real" benefit (in units of goods and
services) provided by the money supply at $t_i$ is defined by
\begin{equation}
l_i=\frac{\lambda_i M_i}{C_i}.
\end{equation}
It follows that we can think of
$\{\lambda_i\}$ as a kind of ``convenience yield" process
associated with the money supply. Rather in the way a country will
obtain a convenience yield (per barrel) from its oil reserves,
which can be expressed on a {\it per capita} basis, likewise an
economy derives a convenience yield (per unit of money) from its
money supply. It is important to note that what matters is the {\it
real benefit} of the money supply, which can be thought of
effectively as a flow of goods and services emanating from the
presence of the money supply. It is possible that the
``wealth" attributable to the face value of the money may in
totality be insignificant. For example, if the money supply
consists of notes issued by the government, and hence
takes the form of government debt, then the {\it per capita}
wealth associated with the face value of the notes is
null, since the representative agent is also responsible
(ultimately) for a share of the government debt. Nevertheless, the
presence of the money supply confers an overall positive flow of
benefit to the agent. If the money supply
consists, say, of gold coins, or units of some other valuable
commodity, then the face value of the money supply will make a
positive contribution to overall wealth, as well as providing a
liquidity benefit.

Our goal is to obtain a consistent structural model for the
pricing kernel $\{\pi_i\}_{i\ge 0}$. We assume that the
representative agent gets utility both from consumption and from
the real benefit of the money supply in the spirit of Sidrauski
1969. Let $U(x,y)$ be a standard bivariate utility function
$U:\R^+\times\R^+\rightarrow\R$, satisfying $U_x>0$, $U_y>0$,
$U_{xx}<0$, $U_{yy}<0$, and $U_{xx}U_{yy}>(U_{xy})^2$. Then the
objective of the representative agent is to maximise an expression
of the form
\begin{equation}
J=\E\left[\sum^{N}_{n=0}\e^{-\gamma t_n}U(k_n,l_n)\right]
\end{equation}
over the time horizon $[t_0,t_1,\ldots,t_N]$, where $\gamma$ is
the appropriate discount rate applicable to delayed gains in
utility. For simplicity of exposition we assume a constant
discount rate. The optimisation problem faced by the agent is
subject to the budget constraint
\begin{equation}
W=\E\left[\sum^N_{n=0}\pi_n(C_n k_n+\lambda_n M_n)\right].
\end{equation}
Here $W$ represent the total {\it per capita} wealth, in nominal
terms, available for consumption related expenditure over the
given time horizon. The agent can maintain a position in money,
and ``consume" the benefit of the money; or the money position can
be liquidated (in part, or in whole) to purchase consumption
goods. In any case, we must include the value of the benefit of
the money supply in the budget. Since the presence of the money supply ``adds value", we
need to recognise this value as a constituent of the budget. The
budget includes also any net initial funds available, together
with the value of expected income (e.g., derivable from labour
or natural resources) over the relevant period.

The fact that the utility depends on the real benefit of the money
supply, whereas the budget depends on the nominal value of the
money supply, leads to a fundamental relationship between the
processes $\{k_i\}$, $\{M_i\}$, $\{C_i\}$, and $\{\lambda_i\}$.
Introducing a Lagrange multiplier $\mu$, after some re-arrangement
we obtain the associated unconstrained optimisation problem, for
which the objective is to maximise the following expression:
\begin{equation}
\E\left[\sum^N_{n=0}\e^{-\gamma
t_n}U(k_n,l_n)-\mu\sum^N_{n=0}\pi_n C_n(k_n+l_n)\right].
\end{equation}
A straightforward argument then shows that the solution for the
optimal policy (if it exists) satisfies the first order conditions
\begin{equation}
U_x(k_n,l_n)=\mu\e^{\gamma t_n}\pi_n C_n\label{U1},
\end{equation}
and
\begin{equation}
U_y(k_n,l_n)=\mu\e^{\gamma t_n}\pi_n C_n\label{U2},
\end{equation}
for each value of $n$ in the relevant time frame, where $\mu$ is
determined by the budget constraint. As a consequence we obtain
the fundamental relation
\begin{equation}\label{fundrelation}
U_x\left(k_n,\lambda_n M_n/C_n\right)=U_y(k_n,\lambda_n M_n/C_n),
\end{equation}
which allows us to eliminate any one of the variables $k_n$,
$M_n$, $\lambda_n$, and $C_n$ in terms of the other three. In this
way, for a given level of consumption, money supply, and liquidity
benefit, we can work out the associated price level. Then by use
of (\ref{U1}), or equivalently (\ref{U2}), we can deduce the form
taken by the nominal pricing kernel, and hence the corresponding
interest rate system. We also obtain thereby an expression for the
``real" pricing kernel $\{\pi_i C_i\}$.

We shall take the view that aggregate consumption, the liquidity
benefit rate, and the money supply level are all determined
exogenously. In particular, in the information-based framework we
take these processes to be adapted to the market filtration, and
hence determined, at any given time, by the values of the
information variables upon which they depend. The theory outlined
above then shows how the values of the real and nominal pricing
kernels can be obtained, at each time, as functions of the
relevant information variables.

It will be useful to have an explicit example in mind, so let us
consider a standard ``log-separable" utility function of the form
\begin{equation}\label{logU}
U(x,y)=A\ln(x)+B\ln(y),
\end{equation}
where $A$ and $B$ are non-negative constants. From the fundamental
relation (\ref{fundrelation}) we immediately obtain
\begin{equation}
\frac{A}{k_n}=\frac{B}{l_n},
\end{equation}
and hence the equality
\begin{equation}
k_n C_n=\frac{A}{B}\lambda_{n}M_n.
\end{equation}
Thus, in the case of log-separable utility we see that the level
of consumption, in nominal terms, is always given by a fixed
proportion of the nominal liquidity benefit obtained from the
money supply. For fixed values of $\lambda_n$ and $k_n$, we
note, for example, that an increase in the money supply leads to
an increase in the price level.

One observes that in the present framework we {\it derive} an
expression for the consumer price index process. This contrasts
with current well-known methodologies for pricing
inflation-linked securities (see, e.g., Hughston 1998, Jarrow
\& Yildirim 2003) where the form of the consumer price index is
specified on an exogenous basis.

The quantity $k_n C_n/M_n$ is commonly referred to as the
``velocity" of money. It measures, roughly speaking, the rate at
which money changes hands, as a percentage of the total money
supply, as a consequence of consumption. Evidently, in the case of
a log-separable utility (\ref{logU}), the velocity has a fixed
ratio to the liquidity benefit. This is a satisfying conclusion,
which shows that even with a relatively simple assumption about
the nature of the utility we are able to obtain an intuitively
natural relation between the velocity of money and the liquidity
benefit. In particular, if liquidity is increased, then a lower
money supply will be required to sustain a given level of nominal
consumption, and hence the velocity will be increased as well. The
situation when the velocity is constant leads to the so-called
``quantity" theory of money, which in the present approach arises
in the case of a representative agent with log-separable utility
and a constant liquidity benefit.

It is interesting to note that the results mentioned so far, in
connection with log-separable utility, are not too sensitive to
the choice of the discount rate $\gamma$, which does not enter
into the fundamental relation (\ref{fundrelation}). On the other
hand, $\gamma$ does enter into the expression for the nominal
pricing kernel; in particular, in the log-separable case we obtain
the following expression for the pricing kernel:
\begin{equation}
\pi_n=\frac{B\e^{-\gamma t_n}}{\mu\lambda_n M_n}.
\end{equation}
Hence, in the log-separable utility theory we can see explicitly
the relation between the nominal money supply and the term
structure of interest rates.

Consider now a contingent claim with the random nominal
payoff $H_j$ at $t_j$. The value of the claim at
$t_0$ in the log-separable utility model is given by the following
formula:
\begin{equation}
H_0=\lambda_0 M_0\e^{-\gamma t_j}\E\left[\frac{H_j}{\lambda_j
M_j}\right].
\end{equation}
One sees two different influences on the value of
$H_0$. First one has the discount factor; but equally
one sees the effect of the money supply. For a given level of the
liquidity benefit (i.e., for constant $\lambda_j$), an increase in the
likely money supply at time $t_j$ will reduce the value of $H_0$.
This example illustrates how market perceptions of the direction
of future monetary policy can affect the valuation of
contingent claims. In particular, the value
of the money supply $M_j$ at $t_j$ will be given as a
function of the best available information at that time concerning
future random factors affecting the economy. The question of how
best to model the money supply process $\{M_i\}$ takes us outside of the realm of pure mathematical finance, and
into the territory of macroeconomics and, ultimately,
political economics. It is interesting to note therefore
that an increase in the liquidity benefit rate has the same practical effect on present valuations as an increase in
the money supply itself.

A striking feature of the separable log-utility example is that
the consumption process does not enter into the valuation formulae
for financial claims. In that case, therefore, one can argue that
inflation is a purely monetary phenomenon, insofar as expectations
affect present valuations. But for more general utility functions
this is not the case. Let us consider, for example, the case of a separable power
utility function, writing
\begin{equation}
U(x,y)=\frac{A}{p}x^p+\frac{B}{q}y^q,
\end{equation}
where $p,q\in(-\infty,1)\setminus\{0\}$. Then a calculation
analogous to the previous one shows that the consumer price index
is given in this situation by
\begin{equation}
C_n=\left(\frac{A}{B}\right)^{1/(1-q)}\frac{\lambda_n
M_n}{k_n^{(1-p)/(1-q)}}.
\end{equation}
Thus in the case of power utility the dependence of the price
index on consumption, although always an inverse relation,
depends, on the ratio of the coefficients of relative risk
aversion associated with real consumption and the money supply
benefit. The corresponding expression for the nominal pricing
kernel is
\begin{equation}
\pi_n=\frac{B^{\frac{1}{1-q}}}{A^{\frac{q}{1-q}}}\,\frac{\e^{-\gamma
t_n}\,k_n^{\frac{q}{1-q}(1-p)}}{\mu\lambda_n M_n}.
\end{equation}
It follows that the value $H_0$ at time $t_0$ of a contingent claim
with the random payoff $H_j$ at time $t_j$ is given by the following formula:
\begin{equation}
H_0=\frac{\lambda_0 M_0}{k_0^{q(1-p)/(1-q)}}\,\e^{-\gamma
t_j}\,\E\left[\frac{H_j\,k_j^{q(1-p)/(1-q)}}{\lambda_j M_j}\right].
\end{equation}
In this situation we see that the valuation depends not only on
expectations concerning the money supply, but also on the level of
consumption.

\vskip 15pt \noindent {\bf Acknowledgements}. LPH and AM
acknowledge support from EPSRC grant number GR/S22998/01; AM
thanks the Public Education Authority of the Canton of Bern,
Switzerland, and the UK Universities ORS scheme, for support. We
are grateful to T. Bj\"ork, D. C. Brody, I. R. C. Buckley, H. B\"uhlmann, H. Geman, M.
Hinnerich, T. Ohmoto, J. L. Vega, and M. Zervos for helpful ideas
and stimulating discussions. We thank participants
at the ETH Risk Day 2006 Z\"urich meeting  and at the May 2007 AMaMeF meeting, Bedlewo, Poland for useful comments.

\vskip 15pt \noindent {\bf References}.

\end{document}